\documentclass[aps,prl,twocolumn,showpacs]{revtex4}

\usepackage{hyperref,latexsym}
\usepackage{graphicx,color,pstricks}
\usepackage{epsfig,epsf,bm}

\begin{document}
\title{Quantum phases of a Feshbach-resonant atomic Bose gas in one dimension}
\author{Yu-Wen Lee}
\affiliation{Physics Department, Tunghai University, Taichung,
Taiwan, R.O.C.} \email{ywlee@thu.edu.tw}
\author{Yu-Li Lee}
\affiliation{Physics Department, National Changhua University of
Education, Changhua, Taiwan, R.O.C.} \email{yllee@cc.ncue.edu.tw}
\begin{abstract}
 We study an atomic Bose gas with an $s$-wave Feshbach resonance
 in a one-dimensional optical lattice, with the densities of atoms
 and molecules incommensurate with the lattice. At zero temperature,
 most of the parameter region is occupied by a phase in which the
 superfluid fluctuations of atoms and molecules are the predominant
 ones, due to the phase fluctuations of atoms and molecules being
 locked by a Josephson coupling between them. When the density
 difference between atoms and molecules is commensurate with the
 lattice, two additional phases may exist: the two component Luttinger
 liquid where both the atomic and molecular sectors are gapless, and
 the inter-channel charge density wave where the relative density
 fluctuations between atoms and molecules are frozen at low energy.
\end{abstract}

\maketitle

\paragraph{Introduction}

Trapped dilute cold atomic gases are one of the most exciting
fields in condensed matter physics.\cite{A} An important recent
development in this area is the application of Feshbach
resonances. The energy difference between the molecular state and
the two-atom continuum, known as the detuning $\nu$, can be
experimentally tuned by means of a magnetic field. Therefore, by
sweeping the magnetic field from positive to negative detuning
through the Feshbach resonance, it is actually possible to form
molecules in the atomic gas.\cite{JBA}

Depending on the quantum statistics of atoms, the low temperature
properties of a dilute atomic gas with an $s$-wave Feshbach
resonance will be much different. In the case of fermonic atoms, a
crossover from a superfluid (SF) of the Bardeen-Cooper-Schrieffer
type to the Bose-Einstein condensation (BEC) is
predicted.\cite{OG} On the other hand, a quantum phase transition
can occur for bosonic atoms by changing the value of magnetic
field detuning.\cite{RPW} There, two thermodynamically distinct
phases exist at zero temperature: the ``atomic superfluid" (ASF)
phase with both atomic BEC and molecular BEC and the ``molecular
superfluid" (MSF) with molecular BEC only.

\begin{figure}
\begin{center}
 \includegraphics[width=0.8\columnwidth]{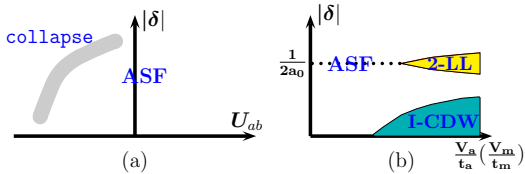}
 \caption{Schematic phase diagram of a Feshbach-resonant atomic
 Bose gas in a one-dimensional optical lattice: (a) on-site
 interactions only and (b) Tonks limit with nearest-neighbor
 repulsions between atoms (molecules) $V_a$ ($V_m$). $\delta$
 denotes the density difference between atoms and molecules.
 $a_0$ is the lattice spacing.}
 \label{1dbf1}
\end{center}
\end{figure}

In the present letter, we study the possible quantum phases for an
atomic Bose gas with an $s$-wave Feshbach resonance in a
one-dimensional optical lattice. It can be described by the
Hamiltonian
\begin{eqnarray}
 H &=& -\sum_i\left(t_mb^{\dagger}_{i+1}b_i+t_aa^{\dagger}_{i+1}
   a_i+{\mathcal H.c.}\right) \nonumber \\
   & & -\sum_i[\mu n_{a,i}+(2\mu -\nu)n_{b,i}] \nonumber \\
   & & +\sum_i\frac{U_a}{2}n_{a,i}(n_{a,i}-1)+\frac{U_m}{2}n_{b,i}
   (n_{b,i}-1) \nonumber \\
   & & +\sum_iU\left(b^{\dagger}_ia_ia_i+{\mathcal H.c.}\right)
   +U_{ab}\sum_in_{a,i}n_{b,i} \ . \label{bfh1}
\end{eqnarray}
Here $n_{a(b),i}$ is the atom (molecule) density operator and
$\mu$ is the chemical potential. This system can also be
considered a kind of binary mixtures with two types of bosons.
However, a main distinction between the present case and the
binary mixture studied previously\cite{D,CH} is that a Josephson
coupling between atoms and molecules, the $U$ term in Eq.
(\ref{bfh1}), is absent in the latter. As we shall see, this will
change the global phase diagram dramatically.

Our main results can be summarized in Fig. \ref{1dbf1}. For
on-site interactions only, i.e. the Hamiltonian given by Eq.
(\ref{bfh1}), the Josephson coupling dominates the low-energy
physics in most of the parameter region, and the resulting phase
exhibits strong superfluid fluctuations in both atomic and
molecular sectors. This is because the phase fluctuations of both
sectors are locked by the Josephson coupling. This phase is, in
fact, a one-dimensional ($1d$) analog of the ASF phase in three
dimensions ($3d$). (But we do not have real condensates here.)
Therefore, it will be referred to as $1d$ ASF. The $1d$ ASF will
collapse for sufficiently strong attractions between atoms and
molecules ($U_{ab}<0$). When the density difference between atoms
and molecules is close to some rational number, depending on the
values of parameters as shown in Fig. \ref{1dbf1} (b), two
additional phases may emerge by including nearest-neighbor
repulsions between atoms (molecules): the two component Luttinger
liquid ($2$-LL) where both the atomic and molecular sectors are
gapless, and the inter-channel charge density wave (I-CDW) where
the relative density fluctuations between atoms and molecules are
frozen at low energy.\cite{foot1}

\paragraph{The continuum theory}

We now outline below the derivation of our results. For
simplicity, we shall consider the case where both the densities of
atoms and molecules, $\rho_a=\langle n_{a,i}\rangle/a_0$ and
$\rho_b=\langle n_{b,i}\rangle/a_0$, respectively, are
incommensurate with the lattice. (Here $a_0$ is the lattice
spacing. $\rho_a$ and $\rho_b$ satisfy the constraint:
$\rho_a+2\rho_b=\rho_0$ where $\rho_0$ is the density of bare
atoms.) That is, we do not consider the possibility of the SF-Mott
insulator transition. Then, in terms of the ``bosonization"
formula\cite{G}: $a_i\sim
e^{-i\sqrt{\pi}\theta_a(x)}\sum_{n=-\infty}^{\infty}e^{i2\pi
n\rho_ax}e^{i\sqrt{4\pi}n\phi_a(x)}$ and $b_i\sim
e^{-i\sqrt{\pi}\theta_m(x)}\sum_{n=-\infty}^{\infty}e^{i2\pi
n\rho_bx}e^{i\sqrt{4\pi}n\phi_m(x)}$, the low-energy physics of
$H$ [Eq. (\ref{bfh1})] can be described by the following effective
Hamiltonian:
\begin{eqnarray}
 H_{eff} &=& \frac{v_a}{2}\int \! \! dx\left[K_a\left(\partial_x
         \theta_a\right)^2+\frac{1}{K_a}\left(\partial_x\phi_a
         \right)^2\right] \nonumber \\
         & & +\frac{v_m}{2}\int \! \! dx\left[K_m\left(\partial_x
         \theta_m\right)^2+\frac{1}{K_m}\left(\partial_x\phi_m
         \right)^2\right] \nonumber \\
         & & +g_1\int \! \! dx\cos{[\sqrt{\pi}(\theta_m-2\theta_a)]}
         \nonumber \\
         & & +g_2\int \! \! dx\partial_x\phi_a\partial_x\phi_m
         \nonumber \\
         & & +g_3\int \! \! dx\cos{[\sqrt{4\pi}(\phi_a-\phi_m)
         +2\pi \delta x]} \ , \label{bfeh1}
\end{eqnarray}
where $\delta =\rho_a-\rho_b$. $v_{a/m}$ and $K_{a/m}$ are sound
velocities and Luttinger liquid (LL) parameters, respectively. In
Eq (\ref{bfeh1}), only those terms which may become the most
relevant in the renormalization group (RG) sense are retained. The
values of $v_{a/m}$, $K_{a/m}$, $g_1$, $g_2$, and $g_3$ depend on
the short-distance physics. In general, they must be extracted
from numerics or experiments. $K_{a/m}\gg 1$ in the weak coupling
regime. On the other hand, $K_{a/m}=1$ in the Tonks limit, i.e.
$U_{a(m)}/t_{a(m)}\rightarrow +\infty$. Therefore, for on-site
interactions only, $1\leq K_{a/m}<+\infty$.\cite{G} The value of
$K_a$ ($K_m$) can be further decreased by including
nearest-neighbor repulsions between atoms (molecules).

The $g_3$ term is a Umklapp process. It can be neglected when
$\delta$ is not close to zero (incommensurate filling). On the
other hand, when $\delta$ is close to zero (commensurate filling),
the $g_3$ term can affect low-energy physics and one may no longer
neglect it in Eq. (\ref{bfeh1}). When $a_0|\delta|$ is close to
some rational number $k/l$ where $k=1,2,\cdots$, $l=2,3,\cdots$,
and $k$ and $l$ are co-prime with one another, other Umklapp
processes must be taken into account, and one must include the
following term in Eq. (\ref{bfeh1}):
\begin{equation}
 \tilde{g}_3\int \! \! dx\cos{[\sqrt{4\pi}l(\phi_a-\phi_m)+2\pi l
 \delta x]} \ . \label{bfeh2}
\end{equation}
We shall focus on Eq. (\ref{bfeh1}) in the following and discuss
the effects of the $\tilde{g}_3$ term later.

\paragraph{Incommensurate filling}

When $\delta$ is incommensurate with the lattice, one may neglect
the $g_3$ term in Eq. (\ref{bfeh1}). To analyze the effects of the
$g_1$ and $g_2$ terms on low-energy physics, we resort to the RG
method. A perturbative calculation up to the one-loop order
already shows that the $g_1$ and $g_2$ terms alone do not form a
closed operator algebra in the sense of operator product expansion
(OPE). One must include the term
$\partial_x\theta_a\partial_x\theta_m$ in $H_{eff}$ [Eq.
(\ref{bfeh1})]. To simplify the analysis, we consider the case
$v_a=v_m=v_0$. (The effects of velocity anisotropy will be
discussed later.) Further, we rescale $g_1$ by
$a_0^2g_1\rightarrow g_1$. The one-loop RG equations can be
obtained by setting $\lambda_3(l)=0$ in Eqs. (\ref{1dbfrge11}) ---
(\ref{1dbfrge16}). By solving these scaling equations, one may
obtain the following results: Within the weak-coupling region,
i.e. $\pi |g_1|/(2v_0), |g_2|/(2v_0)\ll 1$, the $g_1$ term is
relevant in the regime $D_1<2+\frac{\sqrt{2}\pi |g_1|}{v_0}$,
while it becomes irrelevant for $D_1>2+\frac{\sqrt{2}\pi
|g_1|}{v_0}$, where $D_1$ is the scaling dimension of the $g_1$
term, defined by
\begin{equation}
 D_1=\frac{1}{K_a}+\frac{1}{4K_m} \ . \label{bfd1}
\end{equation}
Therefore, there are two zero-temperature phases. It turns out
that the phase transition between these two phases is of the KT
type. We note that for on-site interactions only, i.e. $1\leq
K_{a/m}<+\infty$, the $g_1$ term is always relevant from Eq.
(\ref{bfd1}).

In the regime where the $g_1$ term is relevant, it is convenient
to define new bosonic fields: $\theta_{\pm}\equiv \theta_a\pm
\frac{1}{2}\theta_m$ and $\phi_{\pm}\equiv \frac{1}{2}\phi_a\pm
\phi_m$. Then, the value of $\langle \theta_-\rangle$ is pinned
and a gap $\Delta_0\sim [\pi |g_1|/(2v_0)]^{2-D_1}$ is opened for
the $\theta_-$ sector. That is, the phase fluctuations of atoms
and molecules are locked by the Josephson coupling. By integrating
out the gapped sector, the low-energy effective Hamiltonian
describing the gapless ($\theta_+$) sector takes the form of LLs:
\begin{equation}
 H_+=\frac{v}{2}\int \! \! dx\left[K\left(\partial_x\theta_+
                \right)^2+\frac{1}{K}\left(\partial_x\phi_+\right)^2
                \right] . \label{bfeh3}
\end{equation}
For $\pi |g_1|/(2v_0), |g_2|/(2v_0)\ll 1$, $v$ and $K$ can be
related to the short distance variables with the help of the
one-loop RG equations, yielding
\begin{eqnarray}
 \frac{v}{v_0} &=& \sqrt{\left(\frac{K_a^*}{4}+K_m^*+2\lambda_4^*\right)
   \! \left(\frac{1}{K_a^*}+\frac{1}{4K_m^*}+2\lambda_2^*\right)} \ ,
   \nonumber \\
 K &=& \sqrt{\frac{K_a^*/4+K_m^*+2\lambda_4^*}
   {1/K_a^*+1/(4K_m^*)+2\lambda_2^*}} \ , \label{1dasf1}
\end{eqnarray}
where $K_a^*=K_a+\frac{1}{2-D_1}$, $K_m^*=K_m+\frac{1}{4(2-D_1)}$,
$\lambda_2^*=\frac{g_2}{2v_0}+\frac{1}{2K_aK_m(2-D_1)}$, and
$\lambda_4^*=-\frac{1}{2(2-D_1)}$. Equation (\ref{1dasf1})
indicates that the $1d$ ASF will become unstable provided that the
inequality is satisfied:
$\frac{1}{K_a^*}+\frac{1}{4K_m^*}<-2\lambda_2^*$. This is because
$v\leq 0$, indicating an instability of this system. In other
words, the system will collapse for sufficiently strong attraction
between atoms and molecules.

The $1d$ ASF can be characterized by the single-particle Green
functions of atoms and molecules:
\begin{eqnarray}
 & & \left\langle \Psi_a(\bm{x})\Psi_a^{\dagger}(0)\right\rangle
     =A_1\left(\frac{a_0}{r}\right)^{\alpha_1}+\cdots \ ,
     \nonumber \\
 & & \left\langle \Psi_m(\bm{x})\Psi_m^{\dagger}(0)\right\rangle
     =A_2\left(\frac{a_0}{r}\right)^{\alpha_2}+\cdots \ ,
     \label{1dbfgf1}
\end{eqnarray}
where $\bm{x}=(\tau ,x)$, $r=\sqrt{(v_+\tau)^2+x^2}$,
$\Psi_a(x)=a_i/\sqrt{a_0}$, $\Psi_m(x)=b_i/\sqrt{a_0}$, $A_{1,2}$
are nonuniversal constants, and
\begin{equation}
 \alpha_2=4\alpha_1=\frac{1}{2K} \ . \label{1dbfgf2}
\end{equation}
We would like to stress that this exact relation between
$\alpha_1$ and $\alpha_2$ [Eq. (\ref{1dbfgf2})] results from the
fact that the phase fluctuations of atoms and molecules are locked
by the Josephson coupling, and is a characteristic of the $1d$
ASF. On the other hand, the density correlation functions of atoms
and molecules at nonzero momenta decay exponentially due to the
presence of the gap $\Delta_0$. Equation (\ref{1dbfgf2}) implies
that both the atomic and molecular sectors exhibit the behavior of
a $1d$ SF as long as $K>1/4$. For $1\leq K_{a/m}<+\infty$, this
condition is satisfied, which can be verified from Eq.
(\ref{1dasf1}). It gives further support to our claim that this
phase is a $1d$ analog of the ASF in $3d$.

In the regime where the $g_1$ term is irrelevant, the excitation
spectrum consists of two branches of gapless excitations with
linear dispersion relations. This phase, which has been thoroughly
discussed in Ref. \onlinecite{CH}, will be referred to as the
$2$-LLs. We just mention that in the present situation this phase
can exist only by including sufficiently strong repulsions between
atoms or molecules. This is because $D_1<2$ for on-site
interactions only.

\paragraph{Commensurate filling}

\begin{figure}
\begin{center}
 \includegraphics[width=0.8\columnwidth]{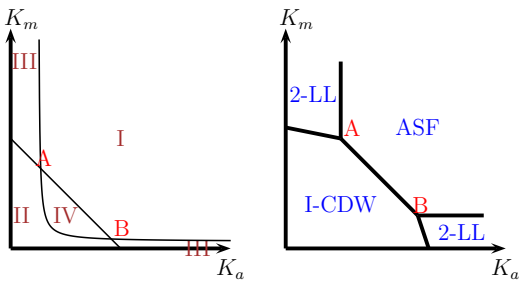}
 \caption{Phase diagram for $\delta =0$: (a)
 suggested by tree-level scaling analysis and (b) predicted by the
 one-loop RG equations. $D_1<2$ and $D_2>2$ in Region I, $D_1>2$ and
 $D_2<2$ in Region II, $D_1,D_2>2$ in Region III, and $D_1,D_2<2$ in
 Region IV. Point $A$ and $B$ in (b) are tricritical points.}
 \label{1dbf2}
\end{center}
\end{figure}

When $\delta$ is close to zero, the $g_3$ term must be retained.
For simplicity, we shall consider $\delta =0$. Let us first define
the scaling dimension of the $g_3$ term:
\begin{equation}
 D_2=K_a+K_m \ . \label{bfd2}
\end{equation}
The $g_1$ ($g_3$) term will be relevant if $D_1<2$ ($D_2<2$).
Accordingly, a tree-level scaling analysis suggests the phase
diagram in the $K_a$-$K_m$ space as shown in Fig. \ref{1dbf2} (a).
$D_1<2$ and $D_2>2$ corresponds to the $1d$ ASF, $D_1>2$ and
$D_2<2$ will be referred to as the I-CDW, and $D_1,D_2>2$
corresponds to the $2$-LL. However, a competition between the two
relevant operators, the $g_1$ and $g_3$ terms, occur when
$D_1,D_2<2$. To determine whether Region IV in Fig. \ref{1dbf2}
(a) corresponds to a new phase or not, we employ the one-loop RG
analysis.

By integrating out the fast modes, the one-loop RG equations are
given by
\begin{eqnarray}
 \frac{dK_a(l)}{dl} &=& 2\left[\lambda^2_1(l)-K^2_a(l)\lambda_3^2(l)
      \right] \ , \label{1dbfrge11} \\
 \frac{dK_m(l)}{dl} &=& \frac{1}{2}\left[\lambda^2_1(l)-4K^2_m(l)
      \lambda_3^2(l)\right] \ , \label{1dbfrge12} \\
 \frac{d\lambda_1(l)}{dl} &=& \! \left[2-\frac{1}{K_a(l)}-\frac{1}
      {4K_m(l)}\right]\lambda_1(l) \nonumber \\
      & & -\frac{\lambda_1(l)\lambda_4(l)}{K_a(l)K_m(l)} \ ,
      \label{1dbfrge13} \\
 \frac{d\lambda_2(l)}{dl} &=& \frac{\lambda_1^2(l)}{K_a(l)K_m(l)}
      -2\lambda_3^2(l) \ , \label{1dbfrge14} \\
 \frac{d\lambda_3(l)}{dl} &=& [2-K_a(l)-K_m(l)]\lambda_3(l) \nonumber
      \\
      & & -2K_a(l)K_m(l)\lambda_2(l)\lambda_3(l) \ , \label{1dbfrge15}
      \\
 \frac{d\lambda_4(l)}{dl} &=& 2K_a(l)K_m(l)\lambda_3^2(l)-\lambda_1^2(l)
      \ , \label{1dbfrge16}
\end{eqnarray}
with the initial values: $K_a(0)=K_a$, $K_m(0)=K_m$,
$\lambda_{1(3)}(0)=\pi g_{1(3)}/(2v_0)$,
$\lambda_2(0)=g_2/(2v_0)$, and $\lambda_4(0)=0$. By solving Eqs.
(\ref{1dbfrge11}) --- (\ref{1dbfrge16}), one may find three kinds
of behaviors of the RG flow of $\lambda_1(l)$ and $\lambda_3(l)$:
(i) $\lambda_1(l)$ flows to strong coupling while $\lambda_3(l)$
flows to zero. This is the $1d$ ASF. (ii) $\lambda_3(l)$ flows to
strong coupling while $\lambda_1(l)$ flows to zero. This is the
I-CDW. (iii) Both $\lambda_1(l)$ and $\lambda_3(l)$ flow to zero.
This is the $2$-LL. Thus, Region IV in Fig. \ref{1dbf2} (a)
shrinks to a transition line between the $1d$ ASF and I-CDW in the
$K_a$-$K_m$ space as shown in Fig. \ref{1dbf2} (b). Both the phase
transition between the $1d$ ASF and $2$ LL and that between the
I-CDW and $2$-LL belong to the KT type. The transition between the
$1d$ ASF and I-CDW is of second order. Further, all these
transition lines coincide at two tricritical points, the point $A$
and $B$ in Fig. \ref{1dbf2} (b). The very reason why there is no
way for both $\lambda_1(l)$ and $\lambda_3(l)$ flowing to strong
coupling simultaneously is that the operators
$\cos{\sqrt{\pi}(\theta_m-2\theta_a)}$ and
$\cos{\sqrt{4\pi}(\phi_a-\phi_m)}$ are exclusive to one another,
that is, the field configurations which minimize one perturbation
term do not minimize the other. The interplay between these two
competing relevant operators then produces a novel quantum phase
transition.

In the I-CDW, it is convenient to define new bosonic fields:
$\tilde{\phi}_+=\frac{1}{2}(\phi_a+\phi_m)$,
$\tilde{\phi}_-=\phi_a-\phi_m$,
$\tilde{\theta}_+=\theta_a+\theta_m$, and
$\tilde{\theta}_-=\frac{1}{2}(\theta_a-\theta_m)$. The
$\tilde{\phi}_+$ and $\tilde{\phi}_-$ fields describe the in-phase
and out-of-phase density fluctuations, respectively. Due to the
relevant perturbation $\cos{\sqrt{4\pi}\tilde{\phi}_-}$, the value
of $\langle \tilde{\phi}_-\rangle$ is pinned and a gap is opened
for the $\tilde{\phi}_-$ sector, while the $\tilde{\phi}_+$ sector
is still gapless. On account of this, both the single-particle
Green functions of atoms and molecules decay exponentially. On the
other hand, the $2\pi \rho_{a/b}$ parts of the density
fluctuations for atoms and molecules are enhanced:
\begin{equation}
 \left.\langle \rho_{a(b)}(x)\rho_{a(b)}(0)\rangle\right|_{2\pi \rho_{a(b)}}
     \sim \left(\frac{a_0}{|x|}\right)^{\gamma} ,
     \label{1dcdwgf1}
\end{equation}
with $\gamma<2$ for $\pi |g_1|/(2v_0), |g_2|/(2v_0)\ll 1$. Here
$\rho_{a/b}(x)=n_{a/b,i}/a_0$.

To understand the nature of the ground state of the I-CDW, a
simple picture can be obtained in the limit of strong
atom-molecule interactions ($g_3$), where the potential energy
(the $g_3$ term) dominates over quantum fluctuations. In this
case, atoms and molecules form a regular lattice (Wigner crystal
of hard-core bosons). For $g_3>0$, the energy of the repulsion
between atoms and molecules
\begin{eqnarray*}
 g_3\cos{(\sqrt{4\pi}\phi_-)}=-g_3\cos{[\sqrt{4\pi}(\phi_a-\phi_m
 +\sqrt{\pi}/2)]} \ ,
\end{eqnarray*}
is minimized by a relative phase shift of $\sqrt{\pi}/2$ between
atoms and molecules, which corresponds to a shift of the atom (or
molecule) lattice by half-a-period.\cite{SMH} Thus, the I-CDW
respects the symmetry of translation by one site, $a_i\rightarrow
a_{i+1}$ and $b_i\rightarrow b_{i+1}$, but spontaneously breaks
the reflection symmetry about the origin, $a_i\rightarrow a_{-i}$
and $b_i\rightarrow b_{-i}$ (or $\phi_{a/m}\rightarrow
-\phi_{a/m}$ and $\theta_{a/m}\rightarrow \theta_{a/m}$).

\paragraph{Experimental signatures}

We suggest a few methods to detect the above results
experimentally. The most important distinction between the 2-LL
and the $1d$ ASF is the behaviors of the density correlators. For
the 2-LLs, the $2\pi \rho_{a(b)}$ part of the density correlators
of atoms (molecules) exhibits power-law decay. However, the
corresponding sector decays exponentially for the $1d$ ASF.
Therefore, a threshold behavior will be observed at the momentum
$k=2\pi \rho_{a(b)}$ in a time-of-flight measurement for atoms
(molecules) in the $1d$ ASF, while such a behavior does not exist
in the 2-LL. The other unique feature of the ASF is that the
ground state is a coherent state formed by hybridizing the atoms
and molecules. Therefore, by suddenly changing the detuning, some
kind of Rabi oscillation will be observed between the atomic and
molecular condensates. Such a change of detuning can be achieved
by applying magnetic field pulses to the ASF.\cite{KH} Although
there is no true condensate in $1d$ due to strong phase
fluctuations, a similar oscillation can also be observed between
the densities of atoms and molecules, which is strongly damped by
the phase fluctuations. Further support to the $1d$ ASF is the
examination of Eq. (\ref{1dbfgf2}), which can be achieved by a
measurement of single-particle Green functions of atoms and
molecules through the absorption line shape.

\paragraph{Discussions}

\begin{figure}
\begin{center}
 \includegraphics[width=0.4\columnwidth]{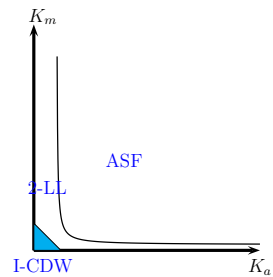}
 \caption{Phase diagram for $a_0|\delta|$ being some rational number.}
 \label{1dbf3}
\end{center}
\end{figure}

Finally, two points should be addressed here. First of all, when
$a_0|\delta|$ is close to some rational number, one must replace
the $g_3$ term in Eq. (\ref{bfeh1}) by the $\tilde{g}_3$ term [Eq.
(\ref{bfeh2})], with the scaling dimension $D_2l$. A tree-level
scaling analysis suggests the phase diagram in the $K_a$-$K_m$
space as shown in Fig. \ref{1dbf3}. We note that there is no
direct transition between the $1d$ ASF and I-CDW. Moreover, the
I-CDW can be reached only with very small values of $K_{a/m}$.
Next, a RG study shows that small velocity anisotropy, i.e.
$v_a\neq v_m$, just shifts the phase boundary, and does not affect
our conclusions. This is because the OPE's between the additional
operators arising from velocity anisotropy and the already
existing operators do not generate themselves.

When this work was finished, we were aware of two recent papers
dealing with a related system --- a Feshbach-resonant atomic Fermi
gas in $1d$.\cite{RO} Their results about the charge sector are
similar to ours for the $1d$ ASF.

\acknowledgments

The work of Y.-W. Lee is supported by the National Science Council
of Taiwan under grant NSC 93-2112-M-029-007. The work of Y.L. Lee
is supported by the National Science Council of Taiwan under grant
NSC 93-2112-M-018-009.



\begin{thebibliography}{99}
 \bibitem{A} M.H. Anderson, J.R. Ensher, M.R. Matthews, C.E. Wieman,
             E.A. Cornell, Science {\bf 269}, 198 (1995); K.B. Davis,
             M. -O. Mewes, M.R. Andrews, N.J. van Druten, D.S. Durfee,
             D.M. Kurn, and W. Ketterle, Phys. Rev. Lett. {\bf 75},
             3969 (1995).
 \bibitem{JBA} S. Jochim, M. Bartenstein, A. Altmeyer, G. Hendl,
             S. Riedl, C. Chin, J. Hecker Denschlag, and R. Grimm,
             Science {\bf 302}, 2101 (2003); M. Greiner, C.A. Regal,
             and D.S. Jin, Nature {\bf 426}, 537 (2003); M.W. Zwierlein,
             C.A. Stan, C.H. Schunck, S.M.F. Raupach, S. Gupta,
             Z. Hadzibabic, and W. Ketterle, Phys. Rev. Lett. {\bf 91},
             250401 (2003).
 \bibitem{OG} Y. Ohashi and A. Griffin, Phys. Rev. Lett. {\bf 89},
              130402 (2002).
 \bibitem{RPW} L. Radzihovsky, J. Park, and P.B. Weichman, Phys. Rev.
             Lett. {\bf 92}, 160402 (2004); M.W.J. Romans, R.A. Duine,
             S. Sachdev, and H.T.C. Stoof, {\it ibid}. {\bf 93}, 020405
             (2004); Y.W. Lee and Y.L. Lee, Phys. Rev. B {\bf 70},
             224506 (2004).
 \bibitem{D} K.K. Das, Phys. Rev. Lett. {\bf 90}, 170403 (2003).
 \bibitem{CH} M.A. Cazalilla and A.F. Ho, Phys. Rev. Lett. {\bf 91},
             150403 (2003); L. Mathey, D.-W. Wang, W. Hofstetter, M.D.
             Lukin, and E. Demler, {\it ibid}. {\bf 93}, 120404 (2004).
 \bibitem{foot1} We emphasize that most of our results can also be
             applied to the case where the Feshbach-resonant atomic gas
             is put in a very elongated trap, such as the toroidal one.
 \bibitem{G} For a review, see, for example, T. Giamarchi, {\it Quantum
             Physics in One Dimension}, (Oxford University Press, Oxford,
             2004).
 \bibitem{SMH} A similar situation also occurs in quantum wires with
             two nearly equivalnet subbands. See, for example, O.A.
             Starykh, D.L. Maslov, W. H\"ausler, and L.I. Glazman in:
             {\it Low-Dimensional Systems: Interaction and Transport
             Properties}, (Springer-Verlag, Berlin, 2000).
 \bibitem{KH} S.J.J.M.F. Kokkelmans and M.J. Holland, Phys. Rev.
             Lett. {\bf 89}, 180401 (2002).
 \bibitem{RO} D.E. Sheehy and L. Radzihovsky, cond-mat/0505681; R.
              Citro and E. Orignac, cond-mat/0505706.
\end{thebibliography}
\end{document}